\documentclass[9pt,a4paper]{extarticle}
\usepackage[a4paper]{geometry}
\usepackage{graphicx} 
\usepackage{tabularx}
\usepackage{xcolor}
\usepackage{float}
\usepackage{xr-hyper}
\usepackage[colorlinks=true, urlcolor=blue, citecolor = blue,]{hyperref}
\usepackage[group-four-digits=true,
            output-decimal-marker={.},
            ]{siunitx}
\usepackage[font=small,labelfont=bf]{subcaption}
\usepackage{caption}
\usepackage[labelfont=bf, figurename=Fig., font={sf,small}]{caption}
\usepackage{natbib}
\setcitestyle{authoryear}
\usepackage{authblk}
\usepackage{enumitem}

\title{\texttt{BikeNodePlanner}: a data-driven decision support tool\\ for bicycle node network planning}

\date{December 2024}

\author[a]{Anastassia Vybornova}
\author[a]{Ane Rahbek Vierø}
\author[b]{Kirsten Krogh Hansen}
\author[a,c]{Michael Szell}

\affil[a]{NEtwoRks, Data, \& Society (NERDS), IT University of Copenhagen, 2300 Copenhagen, Denmark}
\affil[b]{Dansk Kyst- og Naturturisme (DKNT), 9440 Aabybro, Denmark}
\affil[c]{Complexity Science Hub Vienna, 1080 Vienna, Austria}

\begin{document}

\maketitle

\begin{abstract}
\noindent A bicycle node network is a wayfinding system targeted at recreational cyclists, consisting of numbered signposts placed alongside already existing infrastructure . Bicycle node networks are becoming increasingly popular as they encourage sustainable tourism and rural cycling, while also being flexible and cost-effective to implement. However, the lack of a formalized methodology and data-driven tools for the planning of such networks is a hindrance to their adaptation on a larger scale. To address this need, we present the \texttt{BikeNodePlanner}: a fully open-source decision support tool, consisting of modular Python scripts to be run in the free and open-source geographic information system QGIS. The \texttt{BikeNodePlanner} allows the user to evaluate and compare bicycle node network plans through a wide range of metrics, such as land use, proximity to points of interest, and elevation across the network. The \texttt{BikeNodePlanner} provides data-driven decision support for bicycle node network planning, and can hence be of great use for regional planning, cycling tourism, and the promotion of rural cycling.
\end{abstract}


\section{Bicycle node networks: motivation, definition, and implementation}

Recreational cycling is not only an enjoyable way to explore an area, but also helps to popularize cycling among people who do not normally get around by bike, as a nudge to integrate cycling in one's everyday mobility \citep{park_analyzing_2011, boyer_recreational_2018, deenihan_measuring_2013}. Cycling tourism can also decrease mass tourism's burden on environment and climate \citep{kamb_potentials_2021, kim_does_2022}. To promote both recreational cycling and cycling tourism, local governments are increasingly interested in implementing bicycle node networks as a flexible and cost-effective measure. Although bicycle node networks are usually marketed for recreational cycling, they can also boost everyday cycling in rural areas \citep{deenihan_measuring_2013, schmidt_identifying_2024}.

\begin{figure}[ht]
\centering
\includegraphics[width=0.7\textwidth]{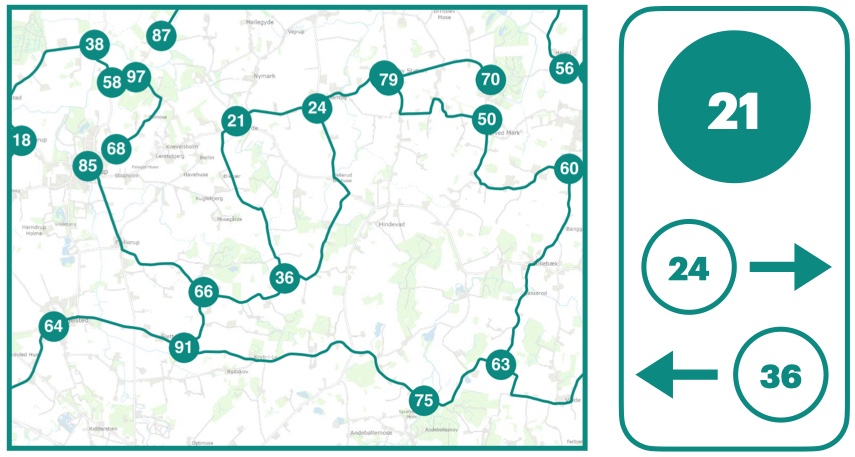}
\caption{\textbf{Left:} Subset of concept bicycle node network. Contrary to a route going straight from A to B, a bicycle node network contains many loops (i.e., possible roundtrips, or ``simple cycles'' in graph theory) and allows for customization of the route. \textbf{Right:} Example of signage for node in the network. The sign points towards the nearest other nodes in the network.}
\label{fig:node_network}
\end{figure}

A bicycle node network is a navigation system tailored to (daytrip) cyclist needs. Seen on a map, the network consists of a set of numbered locations (network nodes) that are connected by roads and paths (network edges) suitable for recreational cycling; and seen from a bicycle, the network consists of signposts placed at each of the nodes (Fig.~\ref{fig:node_network}). The signposts direct the cyclist towards the neighboring nodes on the network. This network layout enables cyclists to plan their routes with maximum flexibility, according to individual needs and preferences. In contrast to many traditional long-haul cycling routes that usually only go from A to B, it allows for a variety of round trips and adjustable trip lengths. 

Implementing a bicycle node network is done by installing signposts for cyclists' wayfinding. It does thus not necessarily require upgrades to the road infrastructure, which makes it potentially much cheaper to implement than e.g.~a network of protected bicycle paths. However, a necessary condition for a good quality bicycle node network is to include only road infrastructure that is suitable for recreational cycling. In addition, the bicycle node network should offer a variation of recreational experiences; provide access to services and amenities along the way; and finally, be safe and well-connected \citep{dknt_metodehandbog_2024}. Hence, to plan such a network for an entire region -- with millions of potential ways to place the nodes, and numerous constraining conditions -- is a complicated logical puzzle. Solving this puzzle manually requires a large amount of humanpower and can be greatly facilitated by data-driven computational methods. 

\section{Previous work on bicycle node network planning}

The first bicycle node network was implemented in Belgium in the 1990s \citep{dknt_studie_2021}. The concept has since been popularized in several other countries, such as the Netherlands, Luxembourg, France, and Germany \citep{nodemapp_cycling_2024, nederland_fietsland_knooppuntroutes_2023, province_de_luxembourg_node--node_2024}, and is currently being implemented in Denmark \citep{caspersen_rekreative_2019, dknt_metodehandbog_2024}. However, there is very little research on bicycle node networks or recreational cycling networks, which is in line with the fact that rural cycling, in contrast to urban cycling, in general remains heavily understudied \citep{mcandrews_motivations_2018, kircher_cycling_2022, scappini_regional_2022, viero_network_2024}. Most literature on cycling tourism and recreational cycling focuses on organizational management, developing attractions and facilities, and only addresses the network design problem through general guidelines \citep{aschauer_guidelines_2021, caspersen_rekreative_2019, dknt_studie_2021, weston_european_2012, wirsenius_cykelleder_2021}. Meanwhile, most literature on data-driven approaches to bicycle network planning is oriented towards adding protected bicycle infrastructure in urban environments for everyday cycling \citep{mauttone_bicycle_2017, caggiani_urban_2019, olmos_data_2020, szell_growing_2022, steinacker_demand-driven_2022, ospina_maximal_2022}. Therefore, up to this date, bicycle node network planning remains a mostly manual process, which requires substantial resources from planners and policy-makers.

Only very few studies have so far aimed to develop methods specifically for recreational cycling or for bicycle node networks. One line of approaches uses mathematical multi-criteria optimization \citep{vansteenwegen_orienteering_2011, cerna_designing_2014, malucelli_designing_2015, giovannini_cycle-tourist_2017, zhu_multi-objective_2022}. Several other studies have documented multi-criteria planning heuristics using desktop GIS \citep{derek_bicycle_2019, scappini_regional_2022}. While these studies are an important first step towards the popularization of bicycle node networks, they are not reproducible in their setup and therefore not directly applicable in planning or policy-making. Moreover, the studies do not consider network structure, which is a crucial factor for bicycle node network design. To the best of our knowledge, no data-driven decision-support tools for bicycle node network planning are available. To address this need, we present the \texttt{BikeNodePlanner}.

\section{Introducing the BikeNodePlanner}

\begin{figure}[ht]
     \centering
     
     \begin{subfigure}[b]{0.3\textwidth}
         \centering
         \includegraphics[width=\textwidth]{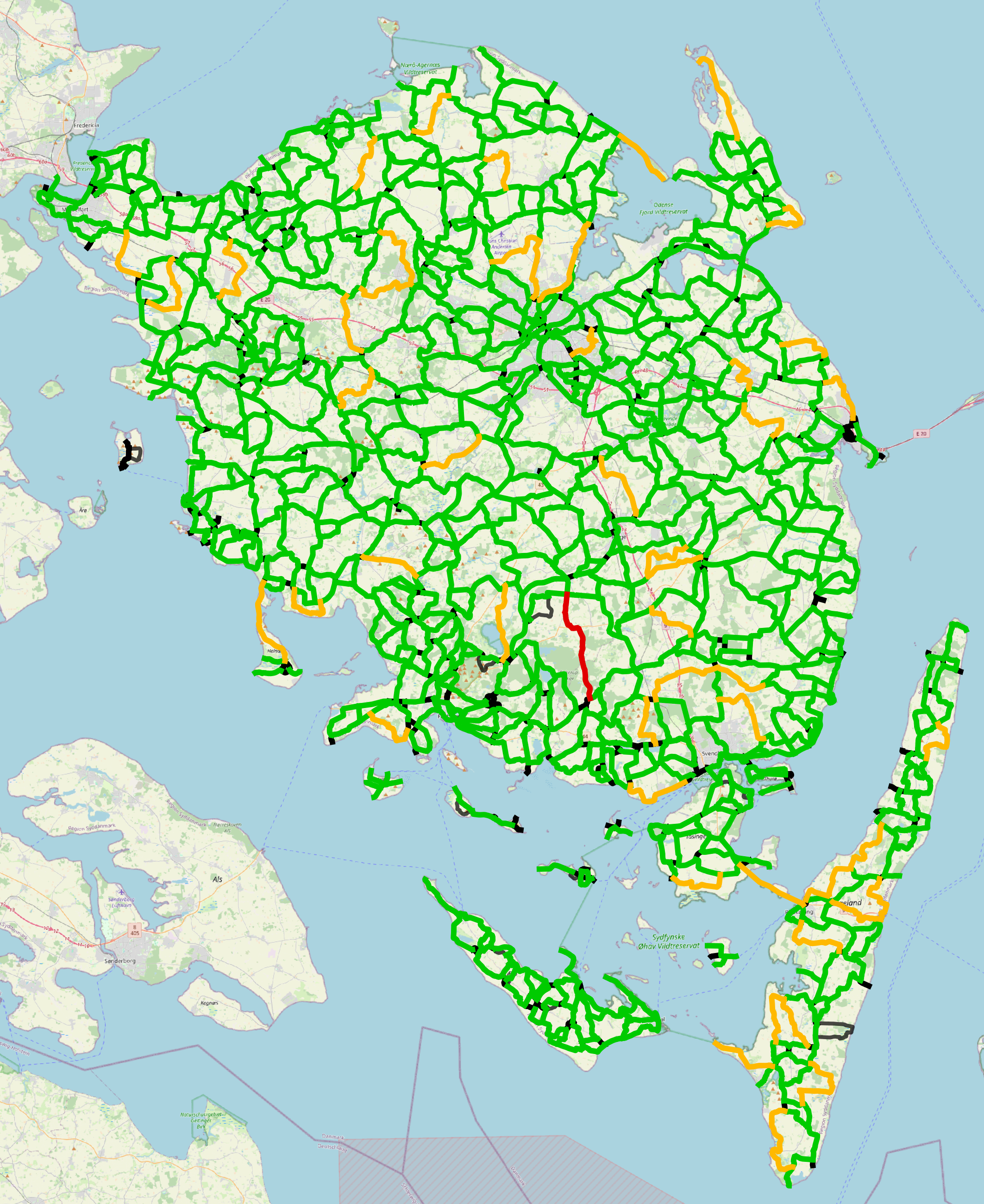}
         \subcaption{Edge length}
     \end{subfigure}
     \hfill
     \begin{subfigure}[b]{0.3\textwidth}
         \centering
         \includegraphics[width=\textwidth]{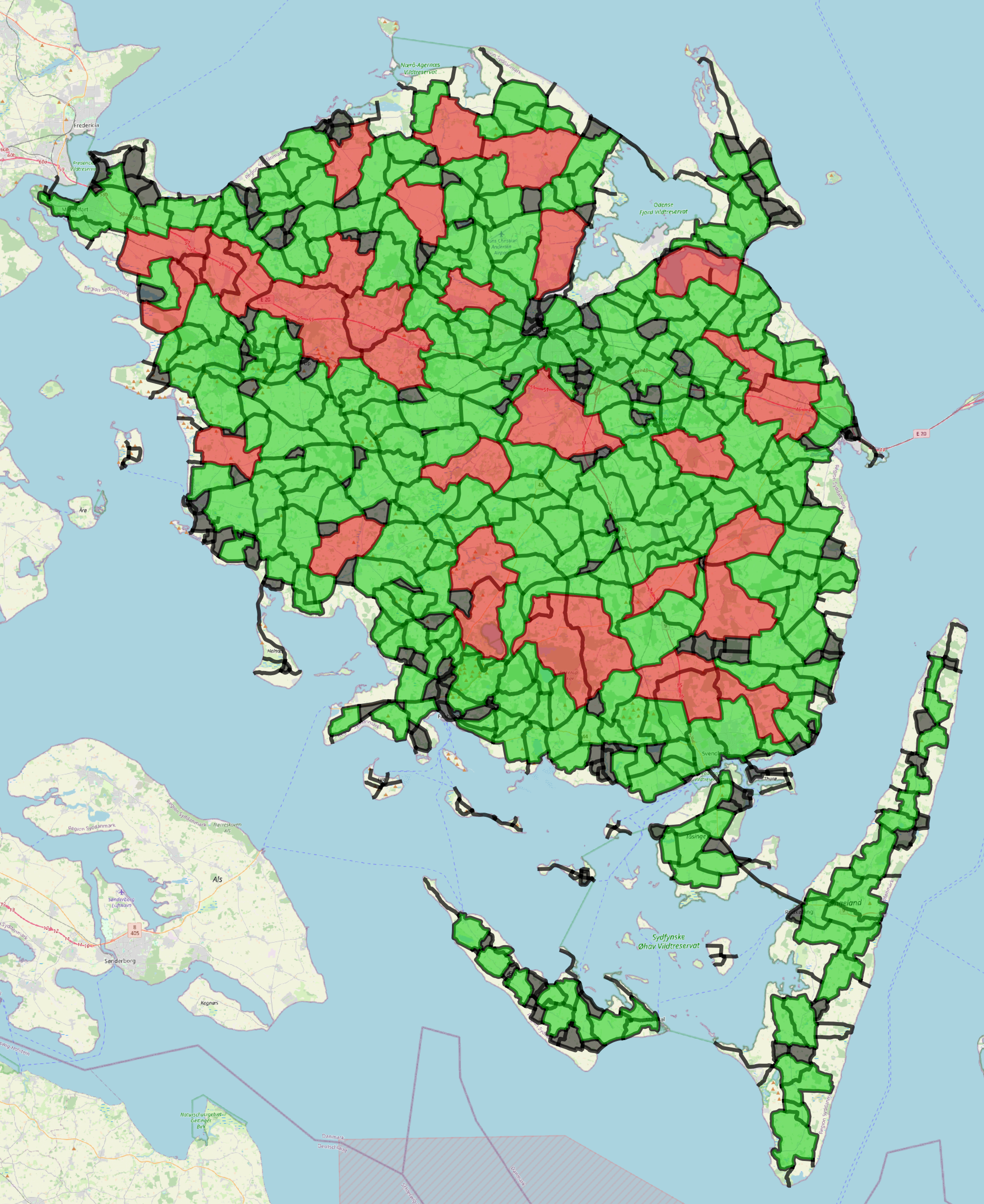}
         \subcaption{Loop length}
     \end{subfigure}
     \hfill
     \begin{subfigure}[b]{0.3\textwidth}
         \centering
         \includegraphics[width=\textwidth]{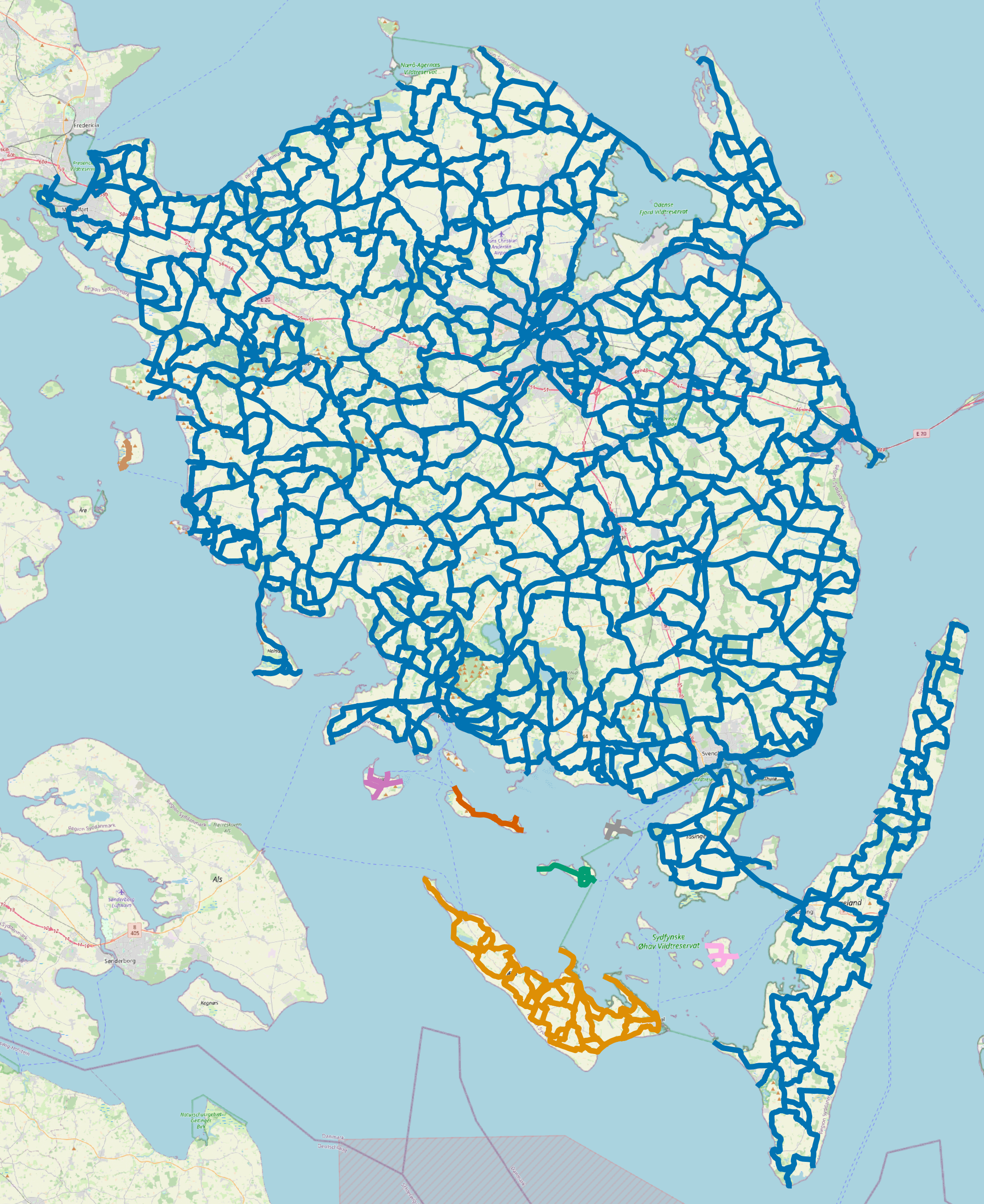}
          \subcaption{Disconnected components}
     \end{subfigure}

\smallskip

     \begin{subfigure}[b]{0.3\textwidth}
         \centering
         \includegraphics[width=\textwidth]{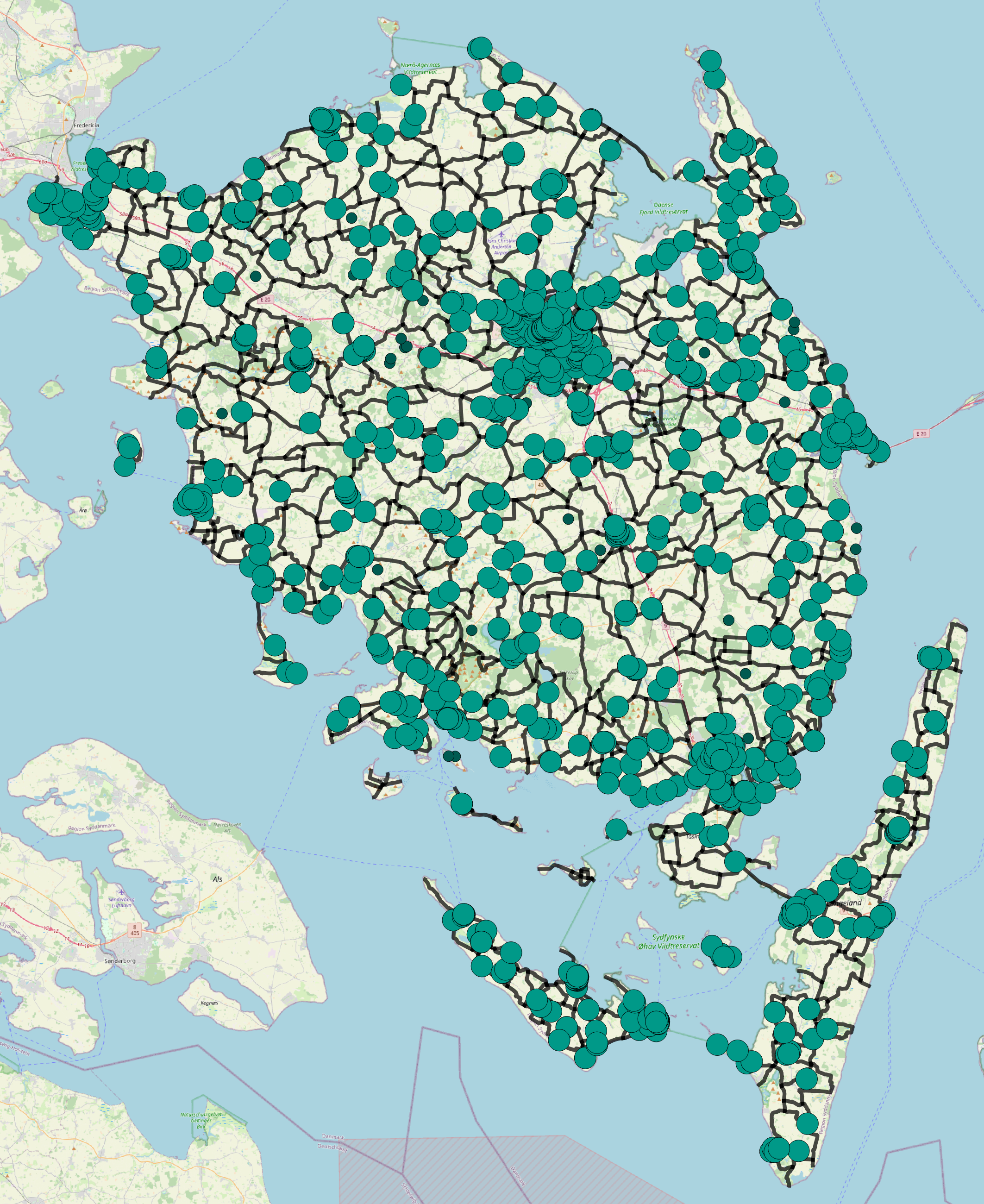}
         \subcaption{Accessibility}
     \end{subfigure}
     \hfill
     \begin{subfigure}[b]{0.3\textwidth}
         \centering
         \includegraphics[width=\textwidth]{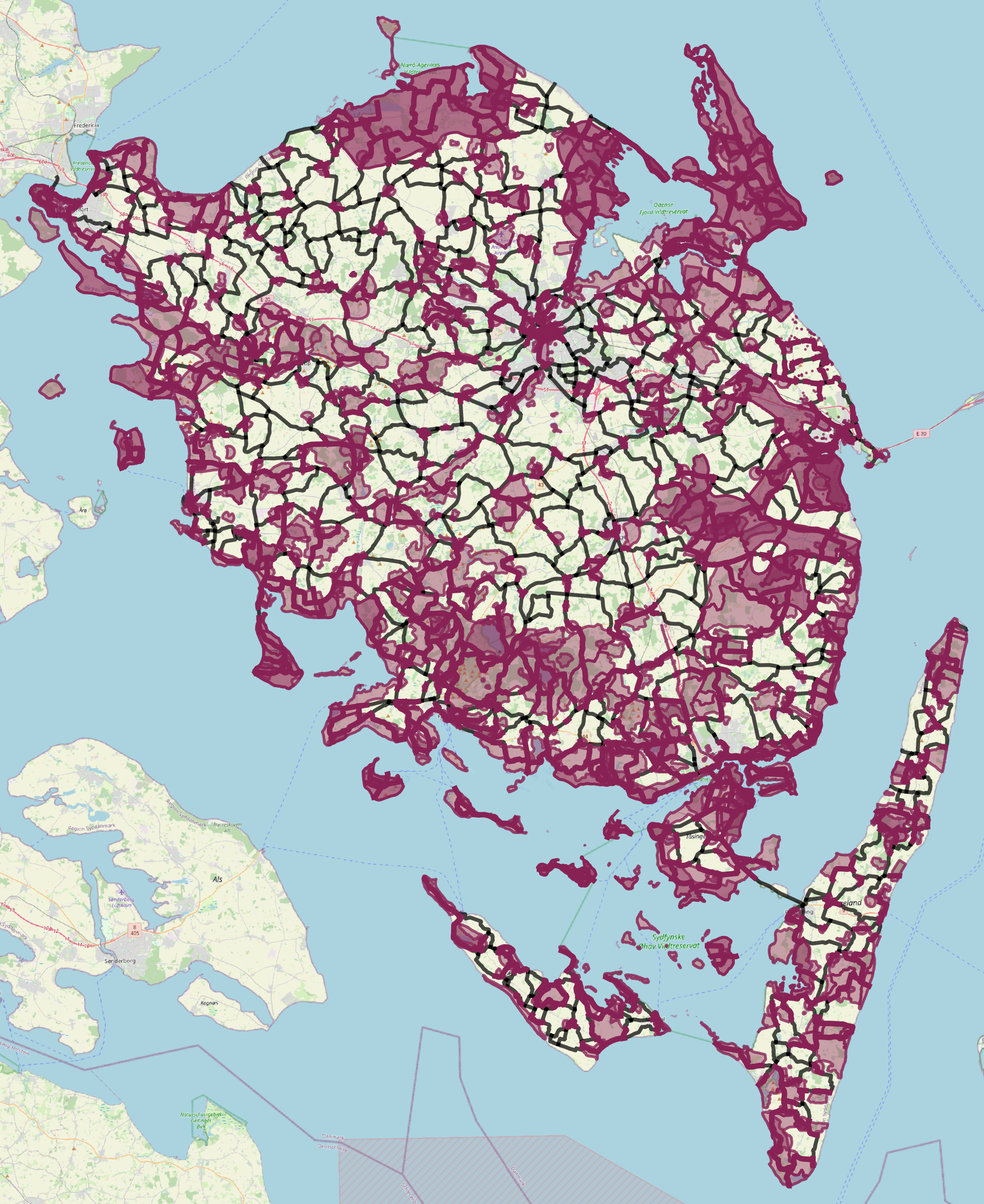}
         \subcaption{Landscape variation}
     \end{subfigure}
     \hfill
     \begin{subfigure}[b]{0.3\textwidth}
         \centering
         \includegraphics[width=\textwidth]{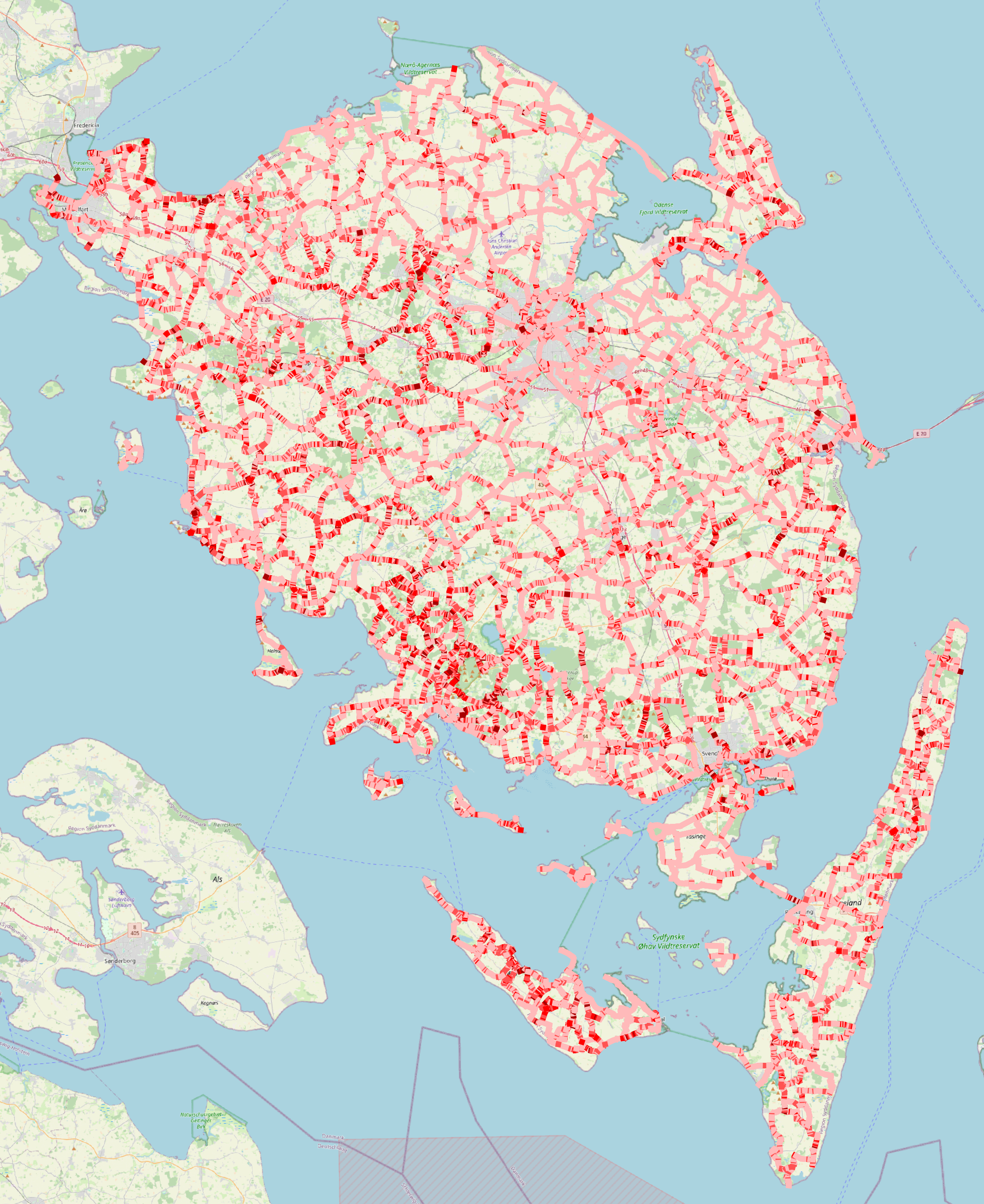}
        \subcaption{Slope}
     \end{subfigure}

        \caption{Overview of \texttt{BikeNodePlanner} features; interactive QGIS legends not shown. \textbf{a)} Classification of network edge length. Black: too short; green: ideal length; yellow: above ideal length; red: too long.  \textbf{b)} Classification of loop lengths. Black: too short; green: ideal length; red: too long. \textbf{c)} Components in the network. Each color represents a disconnected component. \textbf{d)} Network accessibility, point data. Large points represent facilities within reach, small points represent facilities outside of reach, based on the distance threshold.) \textbf{(e)} Landscape variation, polygon data. Highlights where the network goes through areas of cultural interest. \textbf{f)} Network slope. Darker shades of red indicate steeper slopes.}
        \label{fig:qgislayers}
\end{figure}

\begin{table}[ht]
\begin{footnotesize}
\begin{tabularx}{\textwidth}{ 
   >{\centering\arraybackslash \hsize=.5\hsize\linewidth=\hsize}X 
   >{\centering\arraybackslash}X 
   >{\centering\arraybackslash}X 
   >{\centering\arraybackslash}X 
   >{\centering\arraybackslash \hsize=.15\hsize\linewidth=\hsize}X }

Attribute & Relevance & Quantification & \texttt{BikeNodePlanner} & Figure \\
\hline
\\
Edge length & To allow for shorter recreational cycling trips and variation in possible routes  & Ideal length:  $1-5~km$; maximum length:  $10~km $, for dead-ends, maximum length:  $3~km $ & shows edge lengths according to user-defined length thresholds & Fig.~\ref{fig:qgislayers}\textbf{a} \\ \\

Loop length & To allow for shorter recreational cycling trips and variation in possible routes & Ideal length for loops (shortest possible roundtrips): $8-20~km $ & shows loop lengths according to user-defined length thresholds &  Fig.~\ref{fig:qgislayers}\textbf{b} \\ \\

Disconnected components & The network should not have any disconnected components & -- & Identifies and shows separately all disconnected components & Fig.~\ref{fig:qgislayers}\textbf{c} \\ \\

Accessibility of facilities & Necessary facilities such as water, restrooms, and places to buy food should be easily accessible from the network & Toilets: every $10~km$; picnic areas: every  $5~km$. Default distance threshold for reachable facilities: $100~m$ & For a user-defined maximum distance, shows all facilities that are within vs.~outside of reach of the network & Fig.~\ref{fig:qgislayers}\textbf{d} \\ \\

Accessibility of services &  The network should be well-connected to services such as camping sites and hotels to ensure easy access to overnight accommodation for people on multi-day trips & Default distance threshold for reachable services: $750~m$ & For a user-defined maximum distance, shows all services that are within vs.~outside of reach of the network & Fig.~\ref{fig:qgislayers}\textbf{d} \\ \\

Variation in points of interest (POIs) & The network should connect to important POIs: tourist destinations and locations of high recreational value & Default distance threshold for reachable POIs: $1500~m$ & For a user-defined maximum distance, shows all points of interest that are within vs.~outside of reach of the network & Fig.~\ref{fig:qgislayers}\textbf{d} \\ \\

Variation in landscape & A guiding principle for the network planning is to ensure as much variation as possible. One element of variation is to route the network through many different types of landscapes and land use. & -- & For a user-defined maximum distance, and for each landscape (polygon) layer, shows all parts of the network that run through vs.~outside of the layer & Fig.~\ref{fig:qgislayers}\textbf{e} \\ \\

Elevation & It is recommended not to include stretches with too steep slopes. In case of steeper slopes, they should be clearly marked when advertising the network. & Slopes should not exceed 6\% & For user-defined elevation thresholds, shows elevation for all network edges, separately highlighting edges that exceed the maximum slope threshold & Fig.~\ref{fig:qgislayers}\textbf{f}

\end{tabularx}
    \caption{Overview of \texttt{BikeNodePlanner} features, following the bicycle node network planning guidelines by DKNT \citep{dknt_metodehandbog_2024}}
    \label{tab01}
\end{footnotesize}
\end{table}

The \texttt{BikeNodePlanner} is a fully open-source, reproducible PyQGIS tool for decision support in bicycle node network planning, designed for users with minimal experience using GIS software. The tool has been developed in collaboration with \textit{Dansk Kyst- og Naturturisme\footnote{Danish Coast and Nature Tourism}} (DKNT) as part of a larger effort to implement a nationwide bicycle node network in Denmark \citep{dknt_bedre_2021}, but can be applied to any study area in the world. The \texttt{BikeNodePlanner} takes as input (1) a design proposal for a bicycle node network (a spatial data set with network edges and nodes); and (2) additional, user-curated geospatial data on the study area (e.g., land use, location of amenities, elevation). Based on this data, the \texttt{BikeNodePlanner} evaluates the design proposal based on best-practice design criteria for bicycle node networks, as synthesized by DKNT in their recently published handbook on bicycle node network planning \citep{dknt_metodehandbog_2024}. The design criteria and the corresponding \texttt{BikeNodePlanner} evaluation steps are summarized in Table~\ref{tab01}. Each design criterion corresponds to one customizable evaluation step in the \texttt{BikeNodePlanner}. The results can be explored interactively in QGIS. The \texttt{BikeNodePlanner} highlights areas where the network might need to be adjusted, allowing planners to verify whether the network proposal fits the guidelines, and highlighting areas where adjustments might be necessary.  In Fig.~\ref{fig:qgislayers}, we illustrate the main features of the \texttt{BikeNodePlanner} by example of results for the Fyn and Islands (Funen) region of Denmark. Fyn and Islands will also be the first larger location in Denmark where the node network will be implemented with signs and facilities. An exploration of these results and their role for decision-making support is given in the Supplementary Information (SI) (Fig.~\ref{fig:si:edgelooplength}-\ref{fig:si:slopes}). A detailed user guide with feature documentation and data specifications is available in the tool's GitHub repository: \url{github.com/anastassiavybornova/bike-node-planner}.

\section{Workflow overview}

\begin{figure}[H]
\centering
\includegraphics[width=0.7\textwidth]{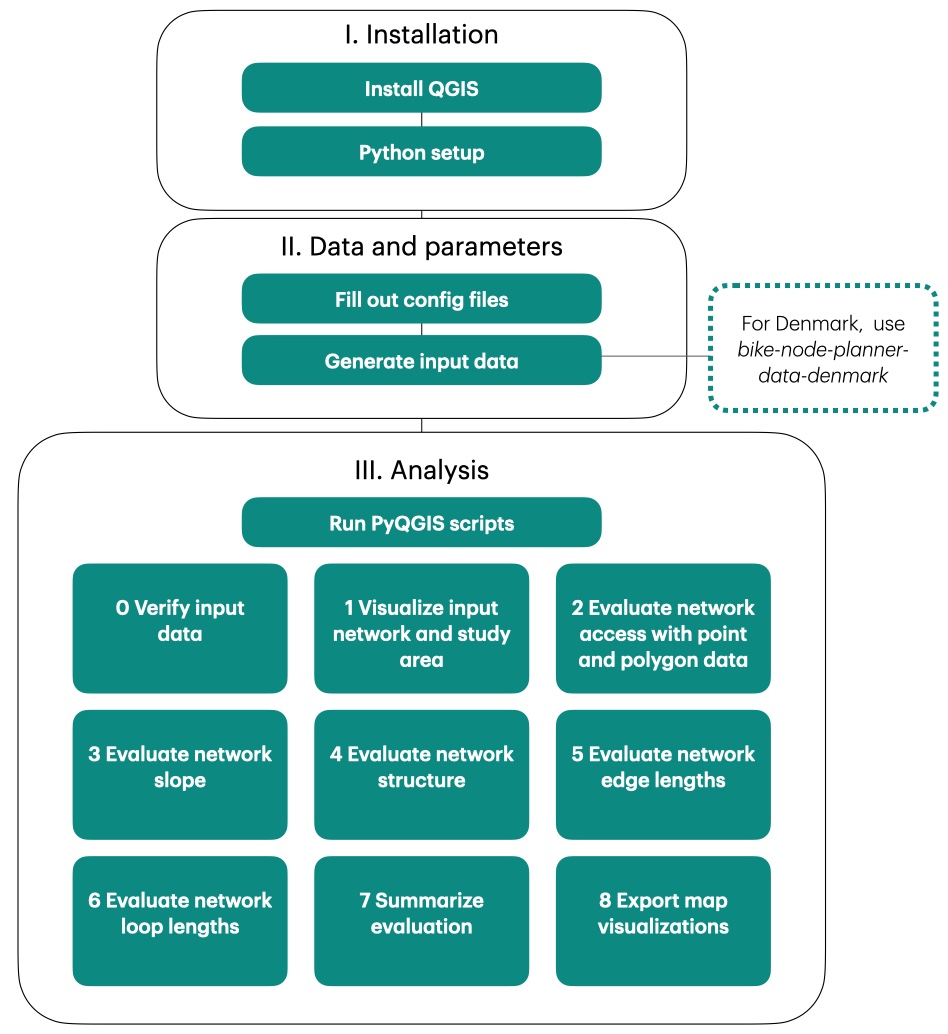}
\caption{The \texttt{BikeNodePlanner} workflow.}
\label{fig:workflow}
\end{figure}

Here, we provide a brief overview of the \texttt{BikeNodePlanner}, as illustrated in Fig.~\ref{fig:workflow}. First, the user installs QGIS and additionally required Python libraries. Next, the evaluation is customized by filling out the configuration files and generating input data\footnote{For Denmark, automated input data generation is available under \url{github.com/anastassiavybornova/bike-node-planner-data-denmark}. For all other study areas, input data must be provided manually; detailed data specifications are available in the \texttt{BikeNodePlanner} documentation (see README Step 3: Prepare your data)}. All input data except the network itself is fully optional, so the \texttt{BikeNodePlanner} can be run independently of evaluation data availability. Now, the user can conduct a step-by-step evaluation of the input network by running the corresponding Python scripts from the QGIS Python console:

\begin{enumerate}[start=0, font=\bfseries]
    \item Verify that input data has been correctly provided
    \item Visualize input network and study area extent
    \item Evaluate network access with point and polygon data, with user-defined distance buffers
    \item Evaluate network slope, with a user-defined classification
    \item Evaluate network structure and display disconnected components
    \item Evaluate network edge lengths, with a user-defined classification
    \item Evaluate network loop lengths, with a user-defined classification
    \item Generate summary plots (incl.~statistics) of all evaluation steps
    \item Export map visualizations of all evaluation steps
\end{enumerate}

\noindent All evaluation layers visualized by the \texttt{BikeNodePlanner} are also stored locally in \texttt{.gpkg} format. The modular nature of the tool, the detailed step-by-step instructions, and the documentation enhance accessibility for users without a programming or GIS software background, while maintaining a high degree of customizability.

\section{Conclusion}

The \texttt{BikeNodePlanner} addresses the need for a decision-support tool for bicycle node network planning that is open-source, customizable, and reproducible. As the first tool of its kind, it incorporates best-practice design criteria for bicycle node network planning \citep{dknt_metodehandbog_2024}, including the structural characteristics of the network. Moreover, the \texttt{BikeNodePlanner} can help identify missing links in existing infrastructure, and therefore can be used to prioritize local improvements to bicycle infrastructure. Although iterative editing (live feedback with continuously updated evaluation results) has not yet been implemented, the tool provides a valuable source of information for planners and policy-makers. Future work -- beyond the localized application of the \texttt{BikeNodePlanner} for selected use cases -- could tackle the challenge of data-driven generation of design proposals for regional bicycle node networks; and focus on increasing network accessibility by public transport. Overall, the \texttt{BikeNodePlanner} is a major first step towards the consolidation of a systematic approach to bicycle node network planning. More broadly, our work contributes to the popularization of bicycle node networks, and to fostering sustainable cycling tourism and rural cycling.

\section{Reproducibility and technical specifications}

The \texttt{BikeNodePlanner} has been developed on MacOS, and tested on both MacOS and Windows. All code is open-source and available under \url{github.com/anastassiavybornova/bike-node-planner}. By following the instructions provided on GitHub for data set preparation, the user can run the \texttt{BikeNodePlanner} workflow for any location. For Denmark, all data and code necessary to produce and preprocess the input data is available under \url{github.com/anastassiavybornova/bike-node-planner-data-denmark}. 

\section{Acknowledgments}

The authors would like to thank Clément Sebastiao, Trivik Verma, and all colleagues from  DKNT, DCT, Folkersma, Septima, NIRAS, Faxe Kommune, and GeoFyn who contributed to this work with helpful feedback, ideas, and comments. We acknowledge support by the Danish Ministry of Transport (grant number: CP21-033).

\clearpage

\begin{footnotesize}
    \bibliography{references}
\end{footnotesize}

\end{document}


\maketitle

Here, we explore \texttt{BikeNodePlanner} results for the Fyn (Funen) region in Denmark. Going through the design criteria listed in Table 1 in the main text, we highlight how the \textbf{BikeNodePlanner} results can be used for decision-making support to evaluate and adjust a bicycle node network design proposal.

\section*{Edge and loop length}

\begin{figure}[ht]
         \centering
         \includegraphics[width=\textwidth]{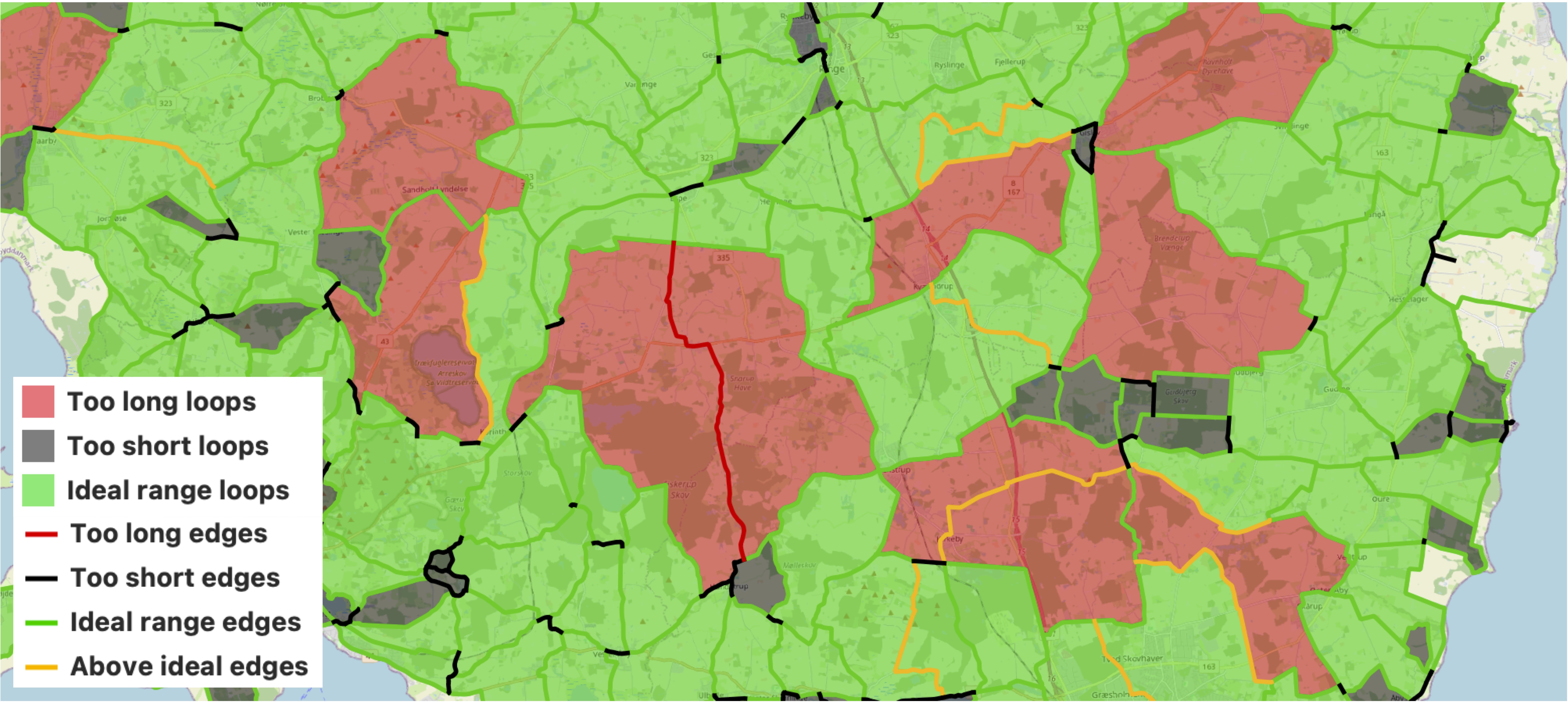}
\caption{\textbf{Edge and loop lengths.} }
\label{fig:si:edgelooplength}
\end{figure}

\noindent To ensure a network that allows flexible trip planning and a wide range of trip distances, the network edges and loop lengths should not be too long \citep{dknt_metodehandbog_2024}. However, too short edges will create a more complicated network and risk information overload by increasing the number of signed nodes.

Network edges are grouped into four classes, with inspiration from the design guidelines from Dansk Kyst og Naturturisme (DKNT) \citep{dknt_metodehandbog_2024}:

\begin{itemize}
    \item Too short: below $1~km$
    \item Ideal range: $1-5~km$
    \item Above ideal: $5-10~km$
    \item Too long: over $10~km$
\end{itemize}

\noindent Network loop lengths (perimeters) are grouped into three classes:

\begin{itemize}
    \item Too short: below $8~km$
    \item Ideal range: $8-20~km$
    \item Too long: over $20~km$
\end{itemize}

\noindent Edges and loops that are classified as too short might be necessary for practical purposes, such as routing through a town; or they might be due to a high density of nodes in the area, which brings the risk of information overload. Edges above ideal length might be unavoidable because of the specific topography of the area; however, if an edge length exceeds the threshold and is classified as too long, it is recommended to introduce an additional node on that edge to reduce its length. The same goes for loops that are classified as too long: introducing additional nodes and/or edges will make the network in the corresponding area more easily navigable, allowing for more flexible route planning.

In the example of Fyn, there are several loops classified as too long; two of them are adjacent and bordered by an edge that is also classified as too long (see Fig.~\ref{fig:si:edgelooplength}). A planner might therefore decide to introduce an additional node on that edge and potentially also introduce additional edges starting at the new node to reduce the size of adjacent loops.

\clearpage

\section*{Disconnected components}

\begin{figure}[ht]
         \centering
         \includegraphics[width=0.7\textwidth]{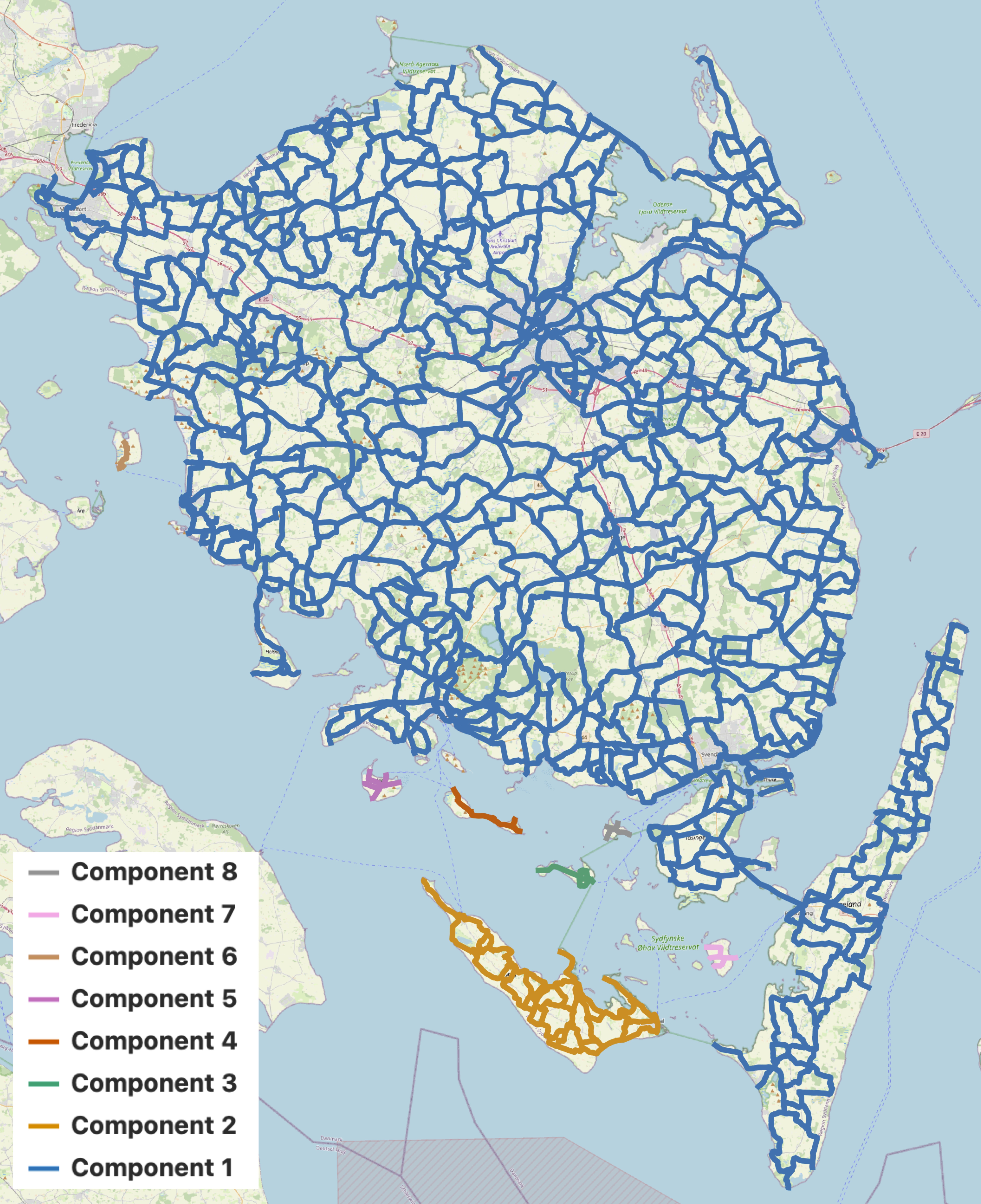}
\caption{\textbf{Disconnected components.} }
\label{fig:si:disconnected-components}
\end{figure}

\noindent Ideally, the network should be well-connected and thus have no disconnected components. In the corresponding layer for Fyn, we can see that the network is naturally split into disconnected components as it spans several islands; on each of the islands, the network consists of one connected component, which is the desired case (see Fig.~\ref{fig:si:disconnected-components}).

\clearpage

\section*{Accessibility}

\begin{figure}[ht]
         \centering
         \includegraphics[width=0.6\textwidth]{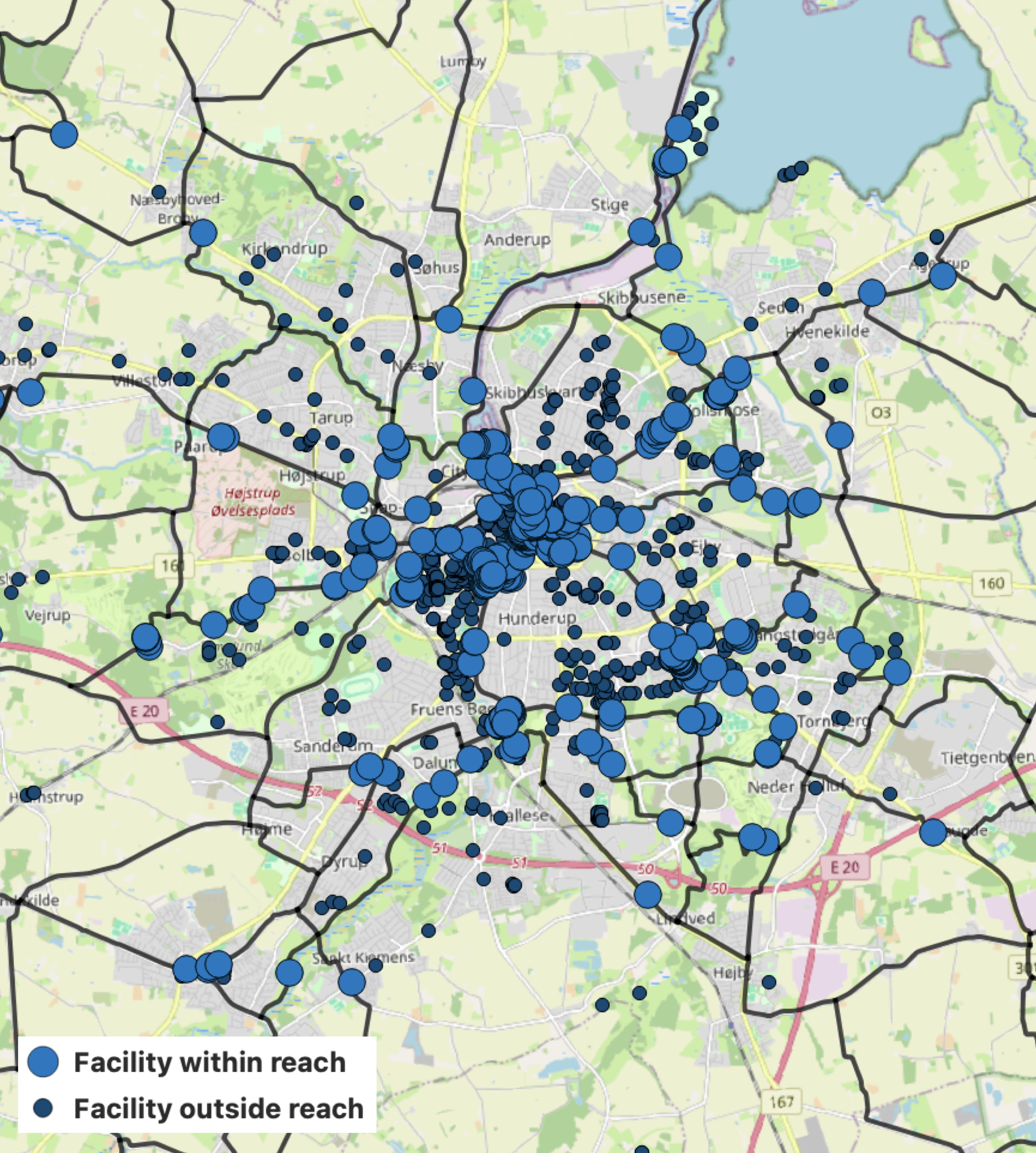}
\caption{\textbf{Accessibility of point data.}}
\label{fig:si:accessibility}
\end{figure}

\noindent To ensure a high recreational value, the network should both connect to important tourist destinations and other points of interest, and provide access to the necessary facilities and amenities. Access to points of interest, services, and facilities are evaluated by classifying network access to corresponding point layers. Points in each point layer are classified as within or outside of reach, depending on the buffer distance provided by the user for the specific layer. For the case of Fyn, we explored three point layers:

\begin{itemize}
    \item Facilities -- this layer combines all amenities that a cyclist might need in immediate proximity along their route, such as water posts, toilets, and grocery shops; the buffer distance is $100~m$.
    \item Service -- this layer combines all locations that are of practical interest and needed at a lesser frequency than facilities, such as restaurants and hotels; the buffer distance is $750~m$.
    \item Points of interest (POIs) -- this layer combines all tourist attractions and sites of interest which make a bigger detour acceptable, such as museums and landmarks; the buffer distance is $1500~m$. 
\end{itemize}

\noindent In addition to the point classification, a heat map for each layer is also provided. This allows to decide, for example, whether a low number of accessible points is due to network routing, or to a generally low point density in the given area. In the example in Fig.~\ref{fig:si:accessibility}, for an area just south of Odense, several facilities within the same loop are classified as out of reach, which might make introducing an additional edge worth considering.

\clearpage

\section*{Landscape variation}

\begin{figure}[ht]
     \begin{subfigure}[b]{0.5\textwidth}
         \centering
         \includegraphics[width=\textwidth]{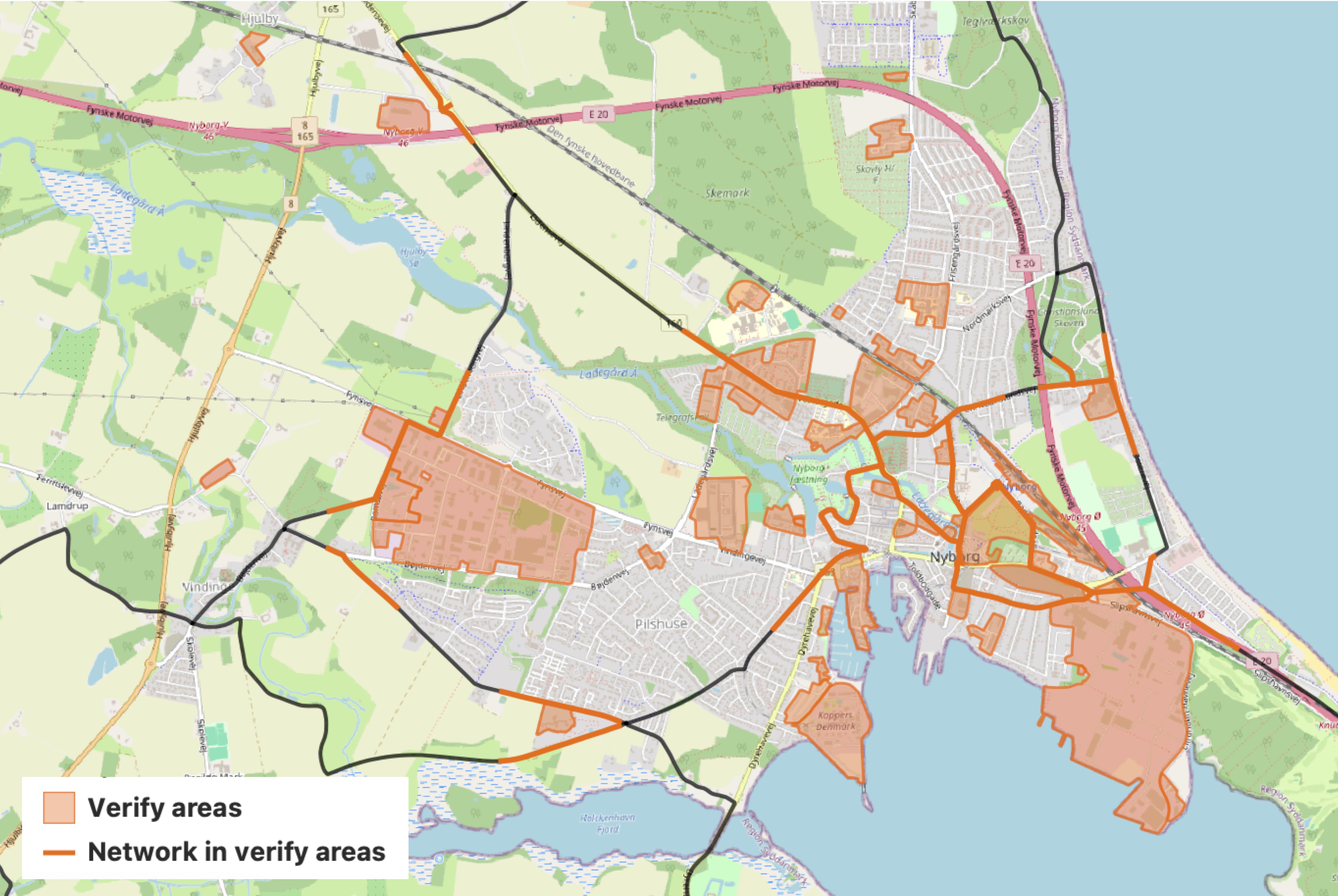}
     \end{subfigure}
     \hfill
     \begin{subfigure}[b]{0.4\textwidth}
         \centering
         \includegraphics[width=\textwidth]{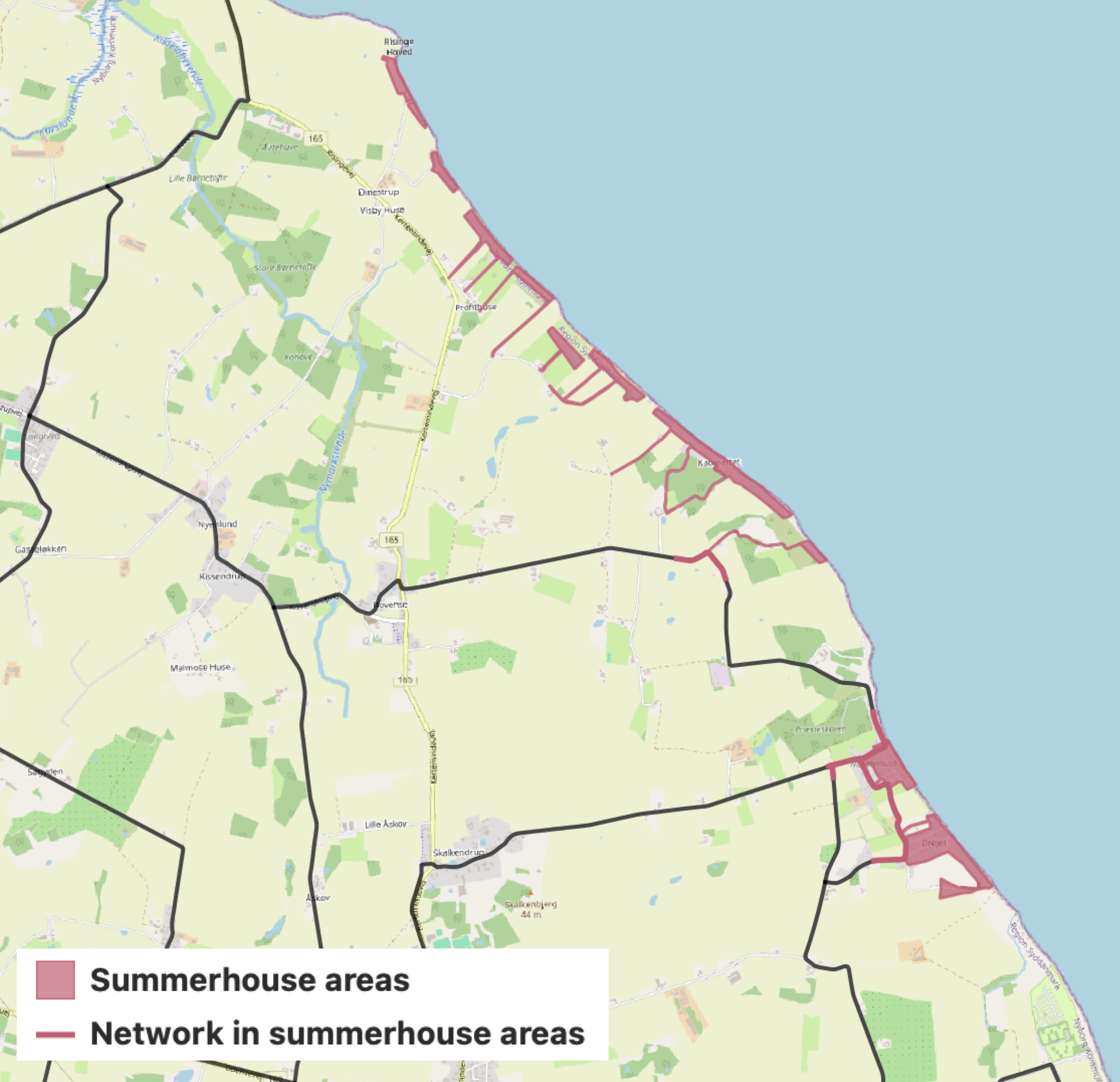}
     \end{subfigure}
\caption{\textbf{Landscape variation.} Left: Zoom-in extract of ``verify'' layer. Right: Zoom-in extract of ``summerhouse'' layer. Network edges running through or in vicinity of each layer are shown in the same color as the layer; rest of edges in black.}
\label{fig:si:poly}
\end{figure}

\noindent A key factor for recreational network planning is to ensure sufficient landscape variation \cite{dknt_metodehandbog_2024}. One way of ensuring this is to analyze the different types of land use surrounding the network. For each polygon layer, the edges that run through that layer are highlighted and summary statistics on network length within vs.~outside each polygon layer are provided. For each layer, edges are buffered by a user-defined buffer distance before evaluation, so that edges running sufficiently close to a given layer are also classified as running through it. For our Fyn example, we summarized available land use data in the following layers:
\begin{itemize}
    \item Nature -- forests, grassland, parks, protected nature zones, lakes, etc.; with a buffer distance of $100~m$
    \item Agriculture -- with a buffer distance of $50~m$
    \item Culture -- all areas of cultural interest, such as historical city centers and prominent landscapes, with a buffer distance of $100~m$
    \item Summerhouse -- all areas with known high density of summer houses (a frequent phenomenon in Denmark, and of particular interest from a cycling tourist perspective); with a buffer distance of $200~m$
    \item Verify -- areas for double-checking whether the network should lead through them or whether it might be advisable to reroute, such as commercial or industrial zones; with a buffer distance of $250~m$
\end{itemize}

\noindent For the use case of Fyn, zooming into the city of Nyborg, we see that several edges run through (or in the vicinity of) areas to verify (see Fig.~\ref{fig:si:poly}, left panel). Based on the local context, a planner can decide whether to partially reroute the network or whether the edges are well placed as is. Zooming into the northeastern part of the island around Kabinettet (see Fig.~\ref{fig:si:poly}, right panel), we see that there is a stretch of summer houses not reached by the network, which makes additional edges worth considering.

\clearpage

\section*{Elevation}

\begin{figure}[ht]
         \centering
         \includegraphics[width=\textwidth]{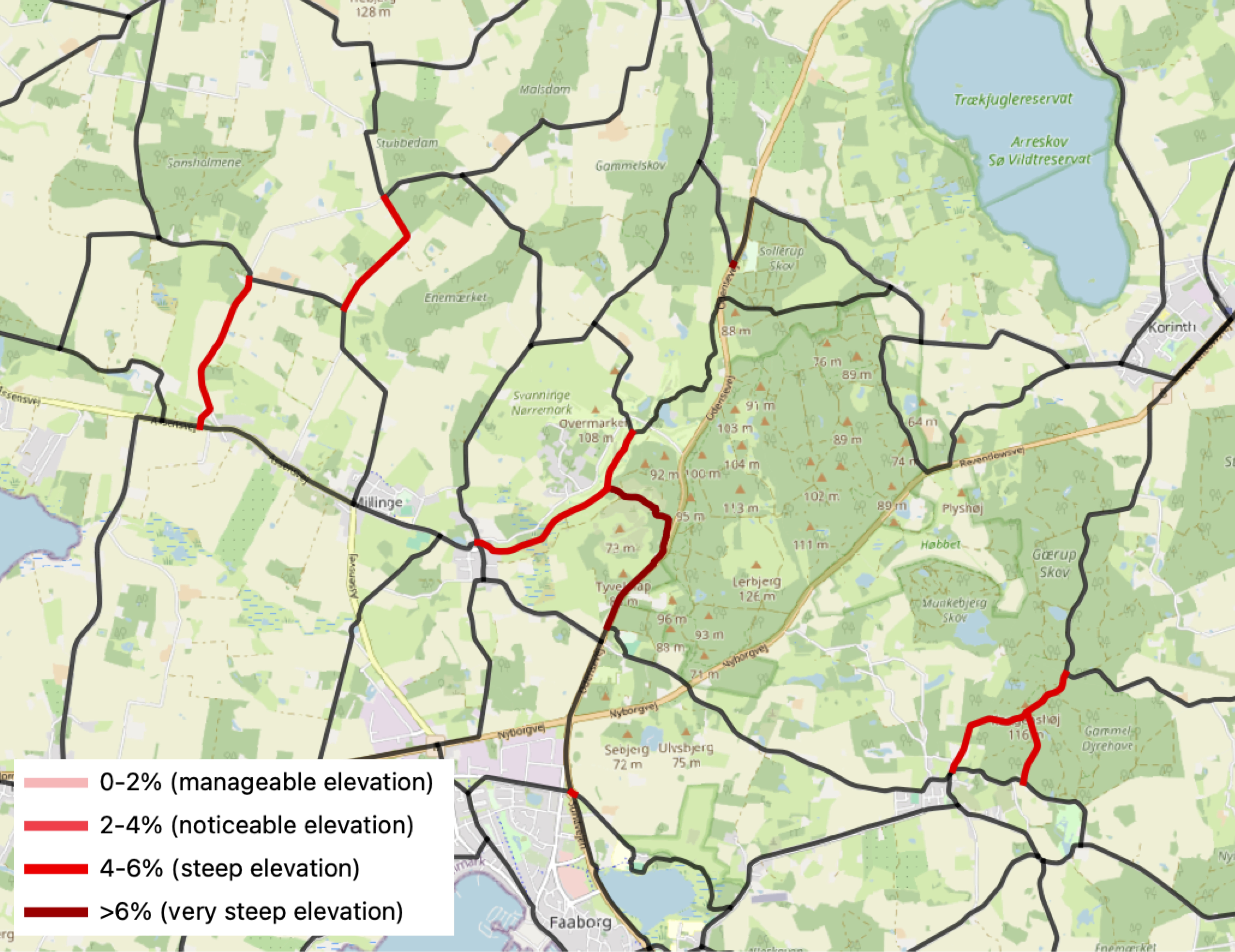}
\caption{\textbf{Average edge slopes.} Only edges of the two steepest classes shown. Edge color indicates average slope: black -- below 4\%; lighter red -- 4-6\%; darker red -- over 6\%}
\label{fig:si:slopes}
\end{figure}

\noindent To make the bicycle node network accessible to all users, regardless of age and physical condition, it is recommended not to include stretches with an elevation exceeding $6\%$ \citep{dknt_metodehandbog_2024}. However, in some regions, this might not be avoidable, depending on the unique landscape of the area. Therefore, the overview of average elevation values across the network is helpful for both planning and public communication. In Fig.~\ref{fig:si:slopes}, we can take a closer look at the area around Knagelbjerg Skov, which contains several steep and very steep edges. A planner might decide to exclude the steep edges from the network -- or only to highlight them separately, so that, e.g., families with children who want to plan a day trip know which areas to avoid. In contrast, cyclists seeking out a mountain bike experience might deliberately choose the same area.

\noindent Edges are grouped into four classes by their average slope:

\begin{itemize}
    \item Manageable: below $2\%$
    \item Noticeable: $2-4\%$
    \item Steep: $4-6\%$
    \item Very steep: over $6\%$
\end{itemize}

\begin{footnotesize}
    \bibliography{references}
\end{footnotesize}